\documentclass[conference]{IEEEtran}
\usepackage{amsmath,amssymb,amsfonts}
\usepackage{algorithmic}
\usepackage{graphicx}
\usepackage{textcomp}
\usepackage{xcolor}
\usepackage{array}
\usepackage{booktabs}
\usepackage{placeins}
\usepackage[numbers]{natbib}
\usepackage{physics}
\usepackage[hidelinks]{hyperref}
\usepackage{ragged2e}
\usepackage{caption}
\def\BibTeX{{\rm B\kern-.05em{\sc i\kern-.025em b}\kern-.08em
    T\kern-.1667em\lower.7ex\hbox{E}\kern-.125emX}}
\begin{document}

\title{Observation of Compressional Acoustic Wave Responses in Cell Culture Media Using a Quartz Crystal Microbalance\\}

\author{
\IEEEauthorblockN{
Hansa Kannan\textsuperscript{1},
Ram Prakash Babu\textsuperscript{2},
Trisha Ghosh\textsuperscript{2},
Arpita Mohapatra\textsuperscript{1},
Mainak Dutta\textsuperscript{1} and Adarsh Ganesan\textsuperscript{2*}
}
\IEEEauthorblockA{
\textsuperscript{1}\textit{Department of Biotechnology, Birla Insitute of Technology and Science, Pilani -- Dubai Campus,}\\
Dubai International Academic City, Dubai, UAE 345055\\
\textsuperscript{2}\textit{Department of Electrical and Electronics Engineering, Birla Insitute of Technology and Science, Pilani -- Dubai Campus,}\\
Dubai International Academic City, Dubai, UAE 345055
\\\textsuperscript{*}Corresponding author: adarsh@dubai.bits-pilani.ac.in
}
}

\maketitle

\begin{abstract}

Quartz Crystal Microbalance (QCM) sensors are widely used to study biological and soft-matter interfaces due to their exceptional sensitivity to mass loading and interfacial mechanical properties. While classical QCM theory assumes predominantly shear-wave coupling into a semi-infinite Newtonian liquid, finite liquid thickness and acoustic reflections give rise to pronounced compressional (longitudinal) wave effects that strongly modulate both resonance frequency and motional resistance. Such compressional acoustic-wave responses should be properly accounted for when sensing in the liquid phase, for instance, when working with cell suspensions. In this work, we systematically investigate compressional-wave responses in cell culture media including DMEM and RPMI-1640 across varying droplet volumes using a 5 MHz AT-cut QCM. Time-resolved measurements are analyzed using four parameters: the time period of compressional acoustic waves ($T_{ca}$), the time associated with a phase shift between resonance frequency and resistance oscillations ($T_{p}$), the peak-to-peak shifts in frequency ($\overline{\Delta f_{pp}}$) and resistance ($\overline{\Delta R_{pp}}$). DMEM and RPMI-1640 both exhibit strong volume-dependent periodic oscillations.
At lower volumes, they exhibit low-frequency oscillations of the time period $\approx 40$ minutes. However, as volume increases, the oscillations gradually evolve into high frequency oscillations of the time period $T_{ca}\approx 5$ minutes. The peak-to-peak shifts ($\overline{\Delta f_{pp}}$) and ($\overline{\Delta R_{pp}}$) are $\approx 100-150 Hz$ and $\approx 40-60\Omega$, respectively. The resonance frequency and resistance oscillations also exhibit a phase shift $T_p$ of $\approx 10$ minutes. These results highlight that compressional-wave artifacts occur even in simple cell culture media, necessitating their explicit consideration when interpreting QCM data in the presence of cells.
\end{abstract}

Quartz Crystal Microbalance (QCM) sensors \cite{Murrieta2022,Carli2024,Chen2025,Wang2025} are widely used to investigate biological interfaces because of their exceptional sensitivity to changes in mass, viscoelasticity, and interfacial mechanical properties \cite{lim2020,jandas2021,alanazi2023,ganesan2016}. Classical QCM theory assumes that the vibrating quartz crystal generates predominantly shear waves that couple into a semi-infinite Newtonian liquid \cite{Sauerbrey1959,voinova1999,johannsmann2015,Fort2022}. However, in some experimental situations, there may also be secondary compressional wave effects due to the longitudinal propagation of acoustic waves into the liquid.

A limited number of studies have reported compressional wave propagation in QCM systems in the past. McKenna et al. \cite{mckenna2001compressional} showed that 2~$\mu$L and 5~$\mu$L microdroplets of water exhibit noticeable compressional waves. This study shows that droplet position and height strongly influence the acoustic response, with good agreement between theoretical predictions and experiments for both static and evaporating microdroplets. Zhang at al. \cite{zhuang2010dynamic} demonstrated these effects in silicone oil droplets with viscosity in a range from $50 cS$ to $103 cS$, resulting in cyclical variations in the frequency and resistance. The effects of interfacial slip and viscoelasticity have been discussed to qualitatively explain the observed positive frequency shift. More recently, Kowarsch et al. \cite{kowarsch2020compressional} report that compressional-wave effects are stronger at the fundamental mode, with the normal motion amplitude ($\sim$1\% of transverse motion) decreasing for higher overtones. Experiments and analytical modeling show that compressional wave and standing-wave effects diminish at higher overtones, consistent with the small-load approximation.

Despite these studies, compressional acoustic wave effects have not yet been reported in biological measurements. This gap is particularly important, as biological assays often rely on liquid-phase environments where additional acoustic phenomena may arise and influence the sensor response. Prior to conducting cell-based experiments, such as cell counting or characterization \cite{Eshun2023,Kowalczyk2023,Ozdagak2023,Palleschi2024,Rogala2025,Kandel2026}, it is therefore essential to first isolate and account for compressional wave effects in the liquid media alone. Neglecting these contributions may lead to misinterpretation of the measured signals, especially when attributing frequency shifts or dissipation changes solely to cellular properties.

In this study, we systematically investigate and demonstrate the presence of compressional acoustic waves in commonly used cell culture media including 
Dulbecco’s Modified Eagle Medium (DMEM) and Roswell Park Memorial Institute (RPMI-1640). By establishing their contribution in the absence of cells, we provide a necessary baseline for accurately interpreting QCM responses in more complex biological environments. This understanding is crucial for improving the reliability and reproducibility of QCM-based measurements in applications involving cell suspensions and other biological system.

In the analysis of QCM operation in liquids, the interaction between the resonator surface and the contacting medium is described through the load impedance \(Z_L\), defined as the ratio of stress to particle velocity. Because the surface motion contains both tangential (shear) and normal (compressional) components, the load impedance naturally separates into two physically distinct contributions. The term \(Z_{L,xzx}\) represents the shear-wave impedance associated with tangential motion, while \(Z_{L,zzz}\) represents the compressional-wave impedance arising from normal motion that launches longitudinal waves into the liquid. Together, these components describe how the liquid loads the resonator within the small-load approximation.

The general tensorial form of the load impedance relates the surface velocity to the resulting stress through

\begin{equation}
\sigma_{\alpha\beta} = Z_{L,\alpha\beta\gamma}\, v_\gamma    
\end{equation}

Here, \(\sigma_{\alpha\beta}\) is the stress tensor, \(v_\gamma\) the velocity component at the surface, and \(Z_{L,\alpha\beta\gamma}\) the third-rank load-impedance tensor. For a thickness-shear resonator, only the components \(Z_{L,xzx}\) and \(Z_{L,zzz}\) are relevant.

\begin{table}[h!] \centering \caption*{\textbf{Table I: Parameter explanation}} 
\begin{tabular}{ll}
\hline
\textbf{Symbol} & \textbf{Description} \\
\hline
$\sigma_{\alpha\beta}$ & Stress tensor component at the resonator surface \\
$v_\gamma$ & Velocity component at the surface \\
$Z_{L,\alpha\beta\gamma}$ & Load-impedance tensor relating stress and velocity \\
$Z_{L,xzx}$ & Shear-wave load impedance \\
$Z_{L,zzz}$ & Compressional-wave load impedance \\
$M_{\mathrm{eff}}$ & Effective mass of the resonator \\
$r_S$ & Position on the resonator surface \\
$v_{\mathrm{ref}}$ & Reference velocity used for normalization \\
$f_0$ & Fundamental resonance frequency \\
$Z_q$ & Shear impedance of quartz \\
$\beta$ & Numerical factor from mode-shape integration \\
$\omega$ & Angular frequency ($2\pi f$) \\
$\rho$ & Density of the liquid \\
$\eta$ & Dynamic viscosity of the liquid \\
$P$ & P-wave (longitudinal) modulus \\
$c_{\mathrm{CW}}$ & Speed of sound in the liquid \\
$k_C$ & Compressional-wave wavenumber \\
$d_{\mathrm{liq}}$ & Thickness of the liquid layer \\
$r_1, r_2$ & Acoustic reflectivities at the cavity boundaries \\
$T$ & Time for liquid interface to move by half a wavelength \\
$t_{\mathrm{off}}$ & Time offset in the impedance oscillation \\
$\alpha$ & Phase associated with reflectivity product \\
$P_{\mathrm{avg}}$ & Time-averaged electrical drive power \\
$R_1$ & Motional resistance of the resonator \\
$I_0$ & Amplitude of the electrical current \\
$v_T$ & Transverse velocity amplitude \\
$d_q$ & Thickness of the quartz plate \\
$A_{\mathrm{eff}}$ & Effective active area of the resonator \\
$e_{26}$ & Piezoelectric stress coefficient \\
$d_{26}$ & Piezoelectric strain coefficient \\
$Q$ & Quality factor of the resonator \\
$n$ & Overtone number \\
$U_T$ & Transverse displacement amplitude \\
\hline
\end{tabular}
\end{table}

The resulting complex frequency shift of the resonator is obtained by integrating the local contributions of shear and normal motion over the crystal surface:

\begin{equation}
\begin{aligned}
\Delta f + i \Delta \Gamma
&=
\frac{i}{4\pi M_{\mathrm{eff}}}
\Bigg[
\int_{\mathrm{Surface}}
\frac{v_x^2(\mathbf{r}_S)}{v_{\mathrm{ref}}^2}
\, Z_{L,xzx}(\mathbf{r}_S)\, \mathrm{d}^2 \mathbf{r}_S \\
&\quad +
\int_{\mathrm{Surface}}
\frac{v_z^2(\mathbf{r}_S)}{v_{\mathrm{ref}}^2}
\, Z_{L,zzz}(\mathbf{r}_S)\, \mathrm{d}^2 \mathbf{r}_S
\Bigg]
\end{aligned}
\end{equation}

In this expression, \(M_{\mathrm{eff}}\) is the effective mass of the resonator, \(r_S\) denotes a point on the surface, and \(v_{\mathrm{ref}}\) is a reference velocity chosen such that the integral of \(v^2/v_{\mathrm{ref}}^2\) equals unity.

When the small-load approximation is valid, the shear and compressional contributions combine linearly, yielding the simplified relation \cite{kowarsch2020compressional}

\begin{equation}
\frac{\Delta f + i\Delta\Gamma}{f_0}
=
\frac{i}{\pi Z_q}
\left(
Z_{L,xzx} + \beta\, Z_{L,zzz}
\right)
\end{equation}

For a Newtonian liquid, the shear-wave impedance is given by

\begin{equation}
Z_{L,xzx} = \sqrt{i\omega\rho\eta},
\end{equation}

with \(\omega = 2\pi f\) the angular frequency, \(\rho\) the liquid density, and \(\eta\) the dynamic viscosity.

If the liquid were semi-infinite with respect to compressional waves, the corresponding compressional-wave impedance would be approximated by

\begin{equation}
Z_{L,zzz} \approx Z_{\mathrm{CW,bulk}}
=
\sqrt{P\rho}
=
c_{\mathrm{CW}}\, \rho
\end{equation}

where \(P\) is the P-wave modulus and \(c_{\mathrm{CW}}\) the speed of sound.

In practice, compressional waves reflect between the resonator surface and the opposite wall of the liquid cell, forming a Fabry–Perot-type cavity. The resulting impedance is described by

\begin{equation}
Z_{L,zzz}
=
Z_{\mathrm{CW,bulk}}
\sum_{n=0}^{\infty}
\left(
r_1 r_2\, e^{2 i k_C d_{\mathrm{liq}}}
\right)^n
=
\frac{Z_{\mathrm{CW,bulk}}}{1 - r_1 r_2\, e^{2 i k_C d_{\mathrm{liq}}}}
\end{equation}

\begin{figure*}
    \centering
    \includegraphics[width=\linewidth]{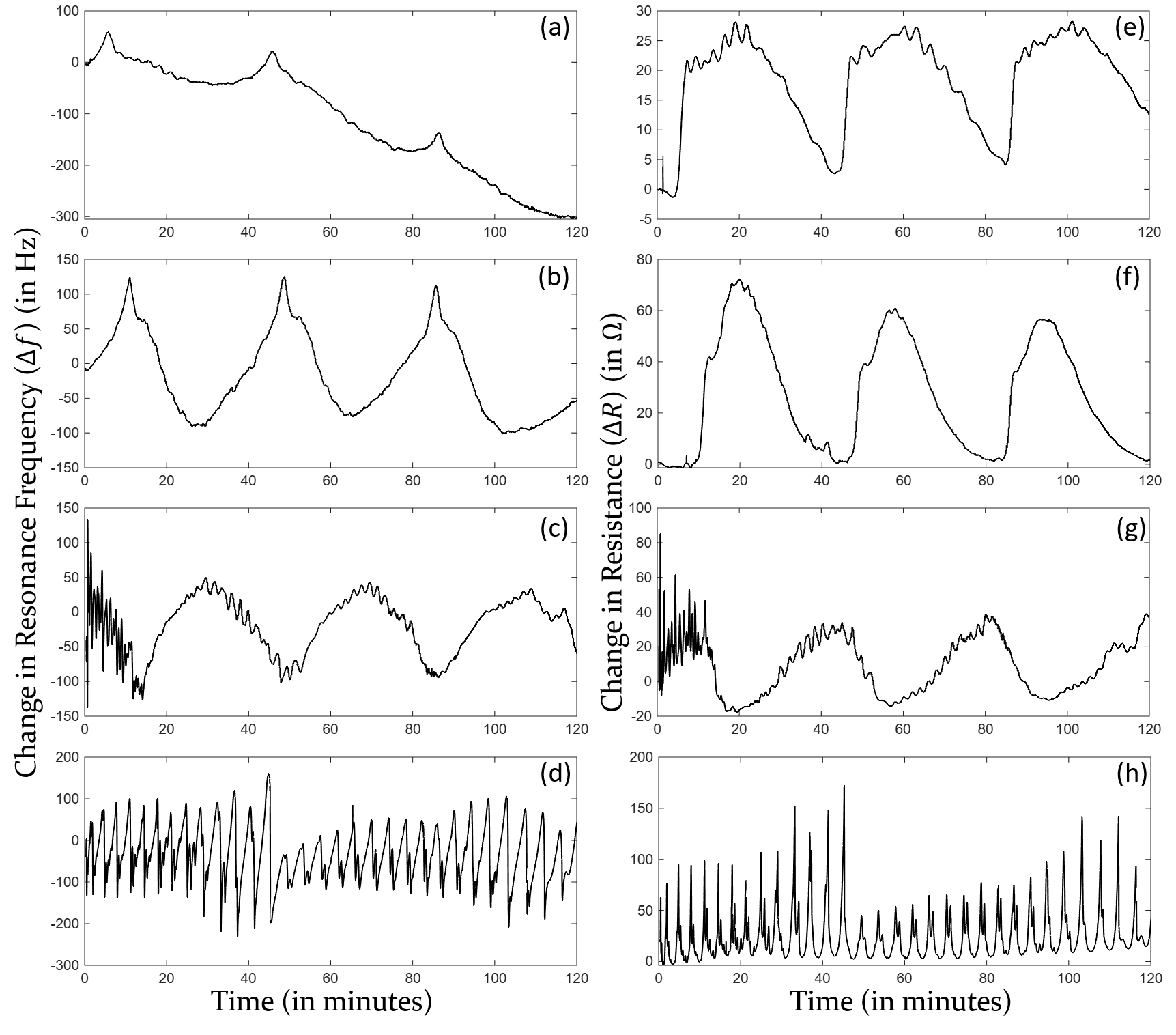}
    \caption{\textbf{QCM measurements of DMEM cell culture media.} Resonance frequency and motional resistance responses at four droplet volumes: (a,e) 400~\textmu L, (b,f) 600~\textmu L, (c,g) 800~\textmu L, and (d,h) 1~mL. Smaller volumes exhibit smooth low-frequency oscillations, whereas larger volumes show pronounced high-frequency fluctuations.}
    \label{fig:1}
\end{figure*}

\begin{figure*}
    \centering
    \includegraphics[width=\linewidth]{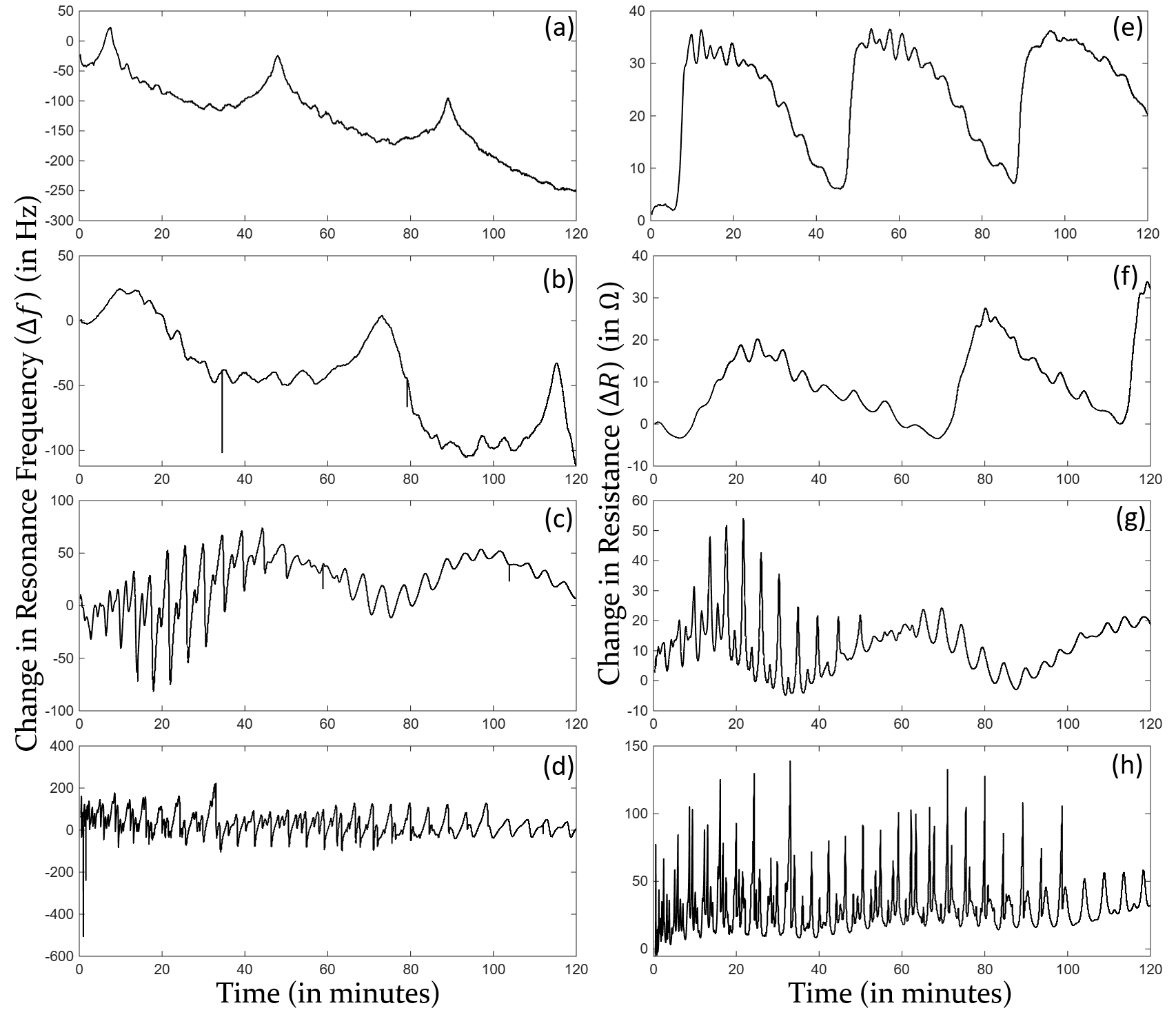}
    \caption{\textbf{QCM measurements of RPMI-1640 cell culture media.} Resonance frequency and motional resistance responses at four droplet volumes: (a,e) 400~\textmu L, (b,f) 600~\textmu L, (c,g) 800~\textmu L, and (d,h) 1~mL. Smaller volumes exhibit smooth low-frequency oscillations, whereas larger volumes show pronounced high-frequency fluctuations.}
    \label{fig:2}
\end{figure*}

As the liquid layer slowly evaporates, the cavity length changes in time. Assuming a constant evaporation rate, the impedance evolves as

\begin{equation}
Z_{L,zzz}(t)
=
\frac{Z_{\mathrm{CW,bulk}}}
{1 - |r_1||r_2|\, e^{i\alpha}\, e^{2\pi i (t - t_{\mathrm{off}})/T}}
\end{equation}

To study the compressional acoustic wave responses, the experimental investigations are conducted using a Quartz Crystal Microbalance (QCM) based measurement system (QCM200 Stanford Research Systems) configured for real-time monitoring of mass variations at the sensor interface. The setup primarily comprised a QCM100 Analog Controller integrated with a QCM25 Crystal Oscillator and interfaced with a computer-based acquisition platform for simultaneous recording of resonance frequency and conductance responses. The system operated with a nominal resonance frequency of 5 MHz employing AT-cut gold-coated quartz crystals, selected due to their superior temperature stability and high sensitivity toward micro-scale mass deposition in liquid and biological environments. Electrical connectivity between the controller and oscillator units is established through a shielded RJ-45 interface cable, while the frequency and conductance outputs are obtained through BNC terminals and relayed to external monitoring and logging instruments.

The sensing element consisted of a gold-plated quartz crystal mounted securely within a dedicated crystal holder assembly. The holder incorporated spring-loaded contact pins to ensure uniform electrical contact with the crystal electrodes and to maintain mechanical stability during liquid loading conditions. Prior to each experimental run, the crystal surface is subjected to a standardized cleaning protocol to eliminate contamination and ensure repeatability of measurements. The sensor is first immersed in 70\% ethanol to remove organic residues and microbial contaminants, followed by rinsing with deionized water to eliminate solvent traces. The crystal is then allowed to air dry under ambient laboratory conditions before being mounted within the holder.

\begin{figure*}
    \centering
    \includegraphics[width=0.8\linewidth]{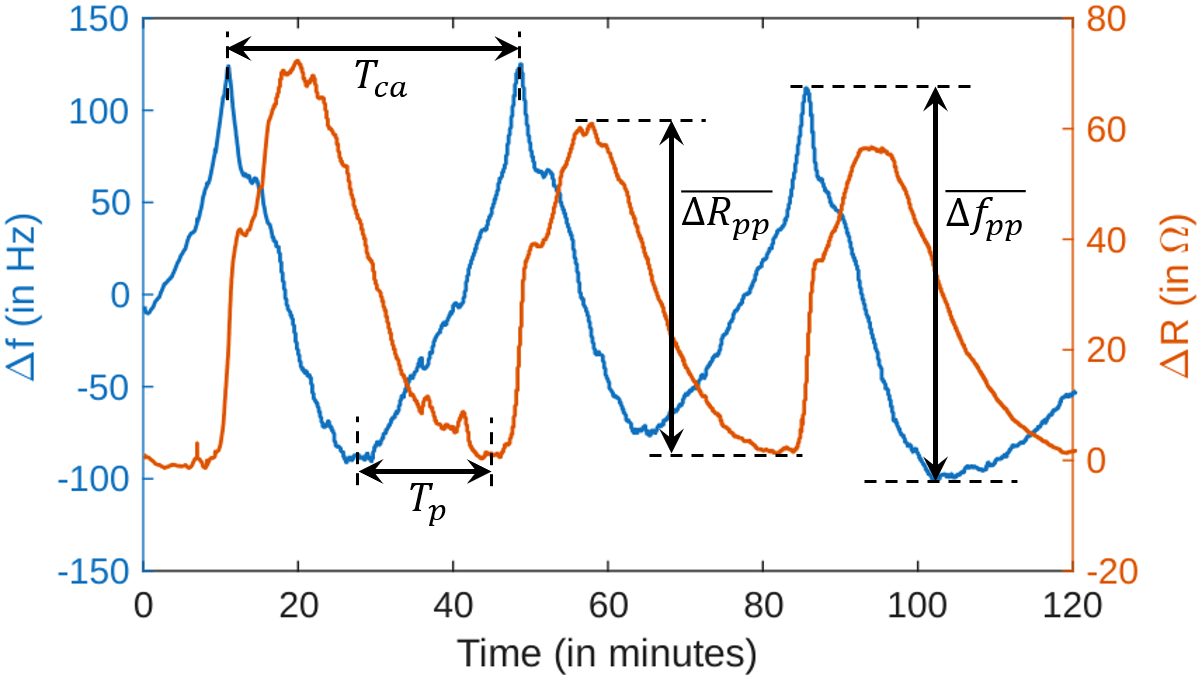}
    \caption{\textbf{Oscillatory parameters} Compressional-wave parameters: time period of compressional acoustic waves ($T_{ca}$), peak-to-peak frequency shift ($\overline{\Delta f_{pp}}$), peak-to-peak resistance shift ($\overline{\Delta R_{pp}}$), and time associated with the phase shift between resonance frequency and resistance oscillations ($T_p$). Note: This plot corresponds to a 600 $\mu$L DMEM droplet.}
    \label{fig:3}
\end{figure*}

\begin{figure*}
    \centering
    \includegraphics[width=\linewidth]{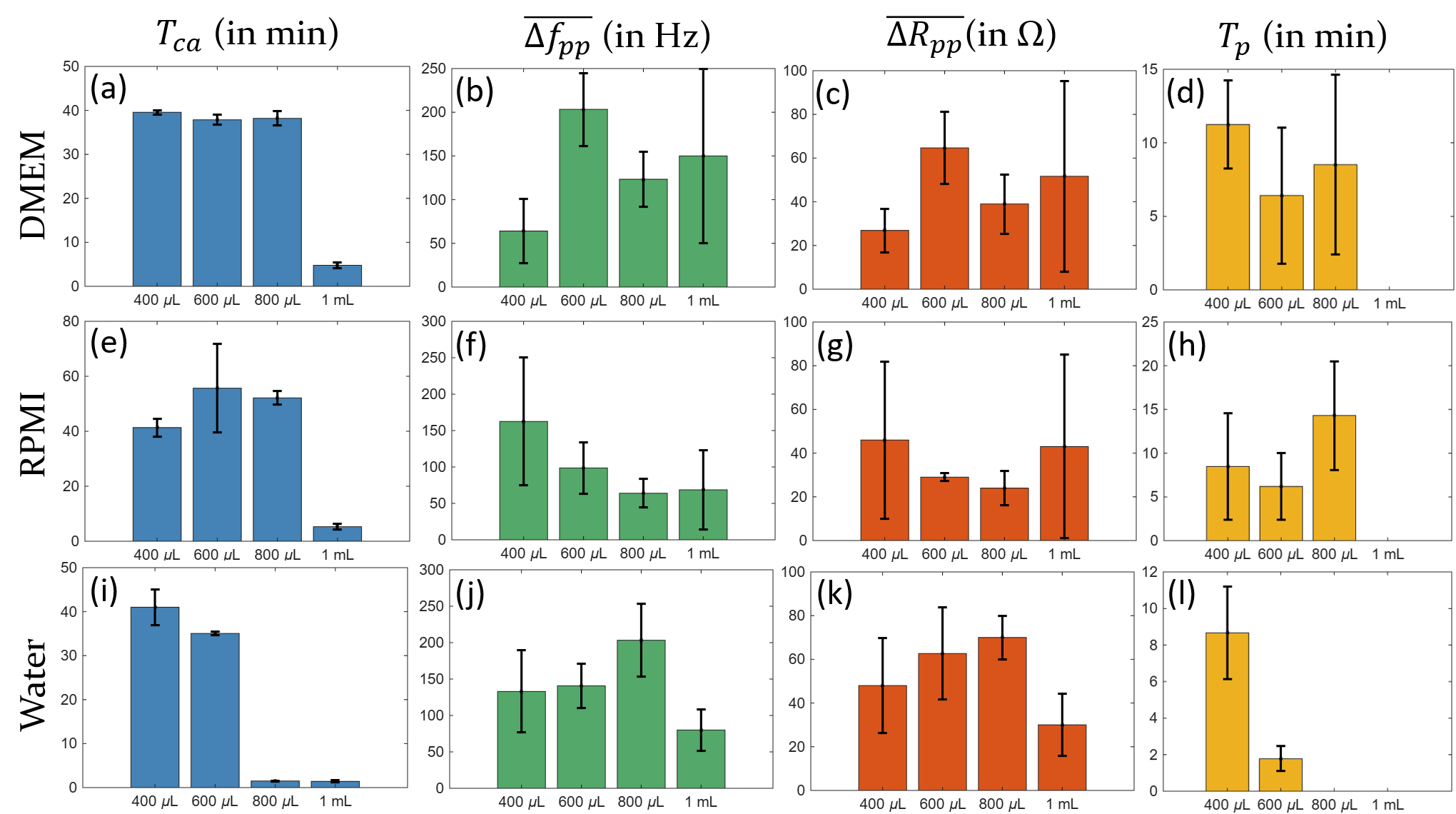}
    \caption{\textbf{Comparison of compressional-wave parameters.} Comparison of $T_{ca}$, $\overline{\Delta f_{pp}}$, $\overline{\Delta R_{pp}}$, and $T_p$, across DMEM, RPMI-1640, and water for four droplet volumes. Here, the error bars indicate the standard deviation of repeated measurements.}
    \label{fig:4}
\end{figure*}

Data acquisition is performed using dedicated QCM interfacing software installed on a laboratory workstation. After sensor mounting and completion of the electrical connections, the system is powered on and allowed to stabilize to obtain baseline frequency and conductance readings. Experimental measurements are initiated through the software interface, enabling continuous logging of sensor responses, which are subsequently exported in spreadsheet format for graphical and analytical processing. Sample introduction is carried out using a calibrated micropipette, wherein the test liquid is carefully dispensed onto the central active surface of the quartz crystal to ensure uniform coverage of the sensing region.

The experimental study included characterization of biologically relevant cell culture media including DMEM (DMEM; high glucose formulation, sigma-Aldrich, cat no. D6429) and RPMI-1640 (Gibco, cat. no. 11875093). These culture media are prepared as a sterile 1X working solution prior to use. To generate complete culture media, the basal culture media are supplemented with 10\%(v/v) fetal bovine serum (FBS) and 1\%(v/v) penicillin-streptomycin solution derived from a 100x stock. All procedures are carried out under aseptic conditions using sterile pipetting techniques within a laminar flow hood. The supplemented media are gently mixed to ensure uniform distribution of components while minimizing floth formation. The prepared media are stored at 2-8 ° C (commonly maintained at 4°C) and utilized within a period of  four weeks. To maintain media integrity and prevent degradation of temperature sensitive components, repeated cycles of warming are avoided.

Upon loading of DMEM or RPMI-1640 sample on QCM surface, the interaction between the deposited mass and the oscillating quartz surface resulted in measurable shifts in resonance frequency along with corresponding variations in motional conductance. These responses are recorded over defined time intervals to capture the behavior of the test media. The experiments are conducted on droplets with four different volumes of DMEM/RPMI-1640: 400 $\mu$L, 600 $\mu$L, 800 $\mu$L and 1 mL. For each volume, at least three identical runs were conducted to ensure statistical significance.

After completion of each measurement cycle, the residual sample was carefully removed, and the crystal underwent the cleaning procedure prior to subsequent testing to prevent cross-contamination. All experiments were conducted under controlled laboratory conditions to minimize environmental disturbances such as temperature fluctuations and mechanical vibrations that could influence frequency stability. The system is allowed adequate warm-up time to achieve oscillator equilibrium, and capacitance compensation procedures were performed before liquid measurements to ensure that the recorded frequency shifts accurately corresponded to mass loading effects at the crystal interface.

Figure~1 shows how the compressional-wave response in DMEM changes as the droplet volume increases. At 400~$\mu$L, the frequency and resistance traces remain smooth and slowly oscillating. Besides this, there is also a continuous decrease in the frequency. However, beyond 600~$\mu$L, the oscillations of resonance frequency and resistance are relatively stable with minimal drift. When the volume reaches 800~$\mu$L and 1~mL, the oscillations become more irregular and high-frequency fluctuations are evidenced. This transition from smooth to highly modulated behavior demonstrates that larger droplets support stronger internal reflections, making the acoustic cavity more active and less stable.

Figure~2 illustrates that RPMI-1640 exhibits a similar volume-dependent oscillatory behavior. 
However, the high-frequency fluctuations become noticeable even at 600~$\mu$L. At 1~mL, the signal becomes highly irregular with strong fluctuations throughout the measurement. These results indicate that RPMI-1640 can generate strong compressional-wave activity even in smaller volumes.

Figures 1(a) and 1(e), and similarly Figures 2(a) and 2(e) indicate that the resonance frequency and resistance oscillations show different patterns as the compressional waves influence mass loading
and dissipation through distinct physical pathways. Figure~3 presents the detailed oscillatory behavior of a 600~$\mu$L DMEM droplet. The $\Delta f$ and $\Delta R$ curves are clearly phase-shifted, confirming that compressional waves modulate mass loading and dissipation differently. The parameters---$T_{ca}$, $\overline{\Delta f_{pp}}$, $\overline{\Delta R_{pp}}$, and $T_p$--- are extracted to quantitatively understand the volume-dependent compressional acoustic-wave behavior.

\begin{itemize}
    \item \textbf{$T_{ca}$ --- Compressional-wave time period} \\
    Represents the duration of one complete oscillation cycle of the compressional wave. It reflects the characteristic timescale of cavity resonance and indicates how rapidly the acoustic field transitions between constructive and destructive interference within the droplet.

    \item \textbf{$\overline{\Delta f_{pp}}$ --- Peak-to-peak frequency modulation amplitude} \\
    Quantifies the total excursion of the frequency shift ($\Delta f$) during each oscillation cycle. A larger $\overline{\Delta f_{pp}}$ signifies stronger modulation of the resonator’s mass-loading response, typically associated with more pronounced compressional-wave activity.

    \item \textbf{$\overline{\Delta R_{pp}}$ --- Peak-to-peak resistance modulation amplitude} \\
    Measures the total excursion of the resistance oscillations ($\Delta R$). This parameter captures the extent of energy dissipation changes induced by the compressional wave.

    \item \textbf{$T_{p}$ --- Time associated with the phase shift} \\
    Denotes the time by which the resistance oscillations lead the resonance frequency oscillations. The observable phase offset between the $\Delta f$ and $\Delta R$ oscillations indicates that compressional waves influence mass loading and dissipation through distinct physical pathways.
\end{itemize}

Figure~4 provides a consolidated comparison of how the key compressional-wave parameters evolve with droplet volume for DMEM, RPMI-1640, and water. In DMEM, all parameters exhibit a strong dependence on volume. At lower volumes, low-frequency oscillatory signatures of $T_{ca}\approx40$ minutes are observed, with $\overline{\Delta f_{pp}}\approx 100-150 Hz$, $\overline{\Delta R_{pp}}\approx 40-60\Omega$ and $T_p\approx10$ minutes. As volume increases, the acoustic cavity supports more pronounced longitudinal-wave interference, resulting in high-frequency fluctuations corresponding to $T_{ca}\approx5$ minutes. The error bars shown in Figure~4 also indicate that the variance of the oscillation time period $T_{ca}$ across multiple identical measurements is much smaller compared to that of the peak-to-peak shifts in resonance frequency $\overline{\Delta f_{pp}}$ and resistance $\overline{\Delta R_{pp}}$.

RPMI-1640 displays a similar trend for all the parameters. Water exhibits the most pronounced compressional-wave behavior in comparison to DMEM and RPMI-1640. Across all volumes, $\overline{\Delta f_{pp}}$ and $\overline{\Delta R_{pp}}$ remain the largest, while $T_p$ and $T_{ca}$ are consistently the shortest. Also, the high-frequency fluctuations of $T_{ca}\approx 2$ minutes are observed at 800 $\mu$L itself. In summary, Figure~4 shows that compressional-wave behavior is governed by the interplay between fluid properties and droplet geometry. Larger droplets, in general, tend to exhibit higher-frequency fluctuations.

This study demonstrates that compressional acoustic waves can significantly influence QCM measurements even when only cell culture media are present on the sensor surface. By analyzing the evolution of resonance frequency and motional resistance for DMEM, RPMI-1640, and water across multiple droplet volumes, we extracted four key parameters—$T_p$, $T_{ca}$, $\overline{\Delta f_{pp}}$, and $\overline{\Delta R_{pp}}$—that quantitatively describe the underlying longitudinal-wave dynamics. Both DMEM and RPMI-1640 exhibit volume-dependent oscillations. At lower volumes, low-frequency oscillations of the time period $\approx 40$ minutes are observed along with weaker high-frequency perturbations. However, as volume increases, high frequency fluctuations become more prominent. 

These observations align with the theoretical framework in which the total load impedance comprises both shear and compressional contributions, and where longitudinal waves reflect between the resonator surface and the liquid–air interface to form a Fabry–Perot-type cavity. The clear dependence of the extracted parameters on both fluid properties and droplet geometry confirms that compressional-wave effects are intrinsic to QCM measurements in bounded liquid layers. For biological and mechanobiological applications, this implies that frequency and dissipation shifts cannot be interpreted solely through shear-driven viscoelastic loading. Instead, longitudinal-wave interference and cavity evolution must be considered, particularly when comparing measurements across different media, volumes, or experimental configurations. The methodology presented here provides a practical route to identify, quantify, and correct for compressional-wave artifacts in QCM-based biosensing \cite{Yilmaz2021, Adel2024, Izzo2026}.

\begin{footnotesize}

\end{footnotesize}

\end{document}